\documentclass[aps,pra,preprint,showpacs,amsmath,amsfonts,amssymb]{revtex4-1}
\usepackage{amsmath,amsfonts,amssymb,color}
\usepackage{amsthm}
\usepackage{leftidx}
\usepackage{graphicx}
\usepackage{xcolor}
\usepackage{dcolumn}
\usepackage{bm}
\usepackage{epstopdf}
\usepackage{epsfig}

\newcommand{\n}{\noindent}

\begin{document}
\title{Generating controllable type-II Weyl points via periodic driving}
\author{Raditya Weda Bomantara}
\affiliation{%
	Department of Physics, National University of Singapore, Singapore 117543
}%
\author{Jiangbin Gong}%
\email{phygj@nus.edu.sg}
\affiliation{%
Department of Physics, National University of Singapore, Singapore 117543
}%
\affiliation{NUS Graduate School for Integrative Science and Engineering, Singapore 117597}
\date{\today}

%%%%%%%%%%%%%%%%%%%% ABSTRACT %%%%%%%%%%%%%%%%%%%%%%%%

\begin{abstract}
Type-II Weyl semimetals are a novel gapless topological phase of matter discovered recently in 2015. Similar to normal (type-I) Weyl semimetals, type-II Weyl semimetals consist of isolated band touching points. However, unlike type-I Weyl semimetals which have a linear energy dispersion around the band touching points forming a three dimensional (3D) Dirac cone, type-II Weyl semimetals have a tilted cone-like structure around the band touching points. This leads to various novel physical properties that are different from type-I Weyl semimetals. In order to study further the properties of type-II Weyl semimetals and perhaps realize them for future applications, generating controllable type-II Weyl semimetals is desirable. In this paper, we propose a way to generate a type-II Weyl semimetal via a generalized Harper model interacting with a harmonic driving field. When the field is treated classically, we find that only type-I Weyl points emerge. However, by treating the field quantum mechanically, some of these type-I Weyl points may turn into type-II Weyl points. Moreover, by tuning the coupling strength, it is possible to control the tilt of the Weyl points and the energy difference between two Weyl points, which makes it possible to generate a pair of mixed Weyl points of type-I and type-II. We also discuss how to physically distinguish  these two types of Weyl points in the framework of our model via the Landau level structures in the presence of an artificial magnetic field. The results are of general interest to quantum optics as well as
ongoing studies of Floquet topological phases.

\end{abstract}
\pacs{42.50.Ct, 42.50.Nn, 03.65.Vf, 05.30.Rt}

%%%%%%%%%%%%%%%%%%%% INTRODUCTION %%%%%%%%%%%%%%%%%%%%%

\maketitle

\section{Introduction}

Since the discovery and realization of topological insulators \cite{SHI, SHI2, SHI3, SHI4}, topological phases of matter have attracted a lot of interests from both theoretical and practical points of view. Topological insulators are characterized by the existence of metallic surface states at the boundaries, which are very robust against small perturbations as long as the topology is preserved. These stable edge states are linked to the topological invariant defining the topological insulator via the bulk-edge correspondence \cite{jack}. As a consequence of their topological properties, topological insulators are potentially useful to generate magnetoelectric effect better than multiferroic materials due to the presence of the axionic term in the electrodynamic Lagrangian \cite{joel}. In addition, Ref.~\cite{maj} shows that by placing a topological insulator next to a superconductor, proximity effect will modify its metallic surface states and turn them into superconducting states. These superconducting states can in turn be used to realize and manipulate Majorana Fermions, which have potential applications in the area of topological quantum computation \cite{collins}.

The interesting properties and potential applications of topological insulators have led to the development of other topological phases. In 2011, Ref.~\cite{Ferarc} discovered a gapless topological phase called Weyl semimetal. Weyl semimetals are characterized by several isolated band touching points in the 3D Brillouin zone, called Weyl points, around which the energy dispersion is linear along any of the quasimomenta forming a 3D Dirac cone. Near these Weyl points, the system can be described by a Weyl Hamiltonian, and the quasiparticle behaves as a relativistic Weyl fermion. Unlike other gapless materials such as Graphene, the Weyl points in Weyl semimetal are very robust against perturbations and cannot be destroyed easily, provided the perturbations respect both translational invariance and charge conservation \cite{Hosur}. Each Weyl point is characterized by a topological charge known as chirality. Under open boundary conditions (OBC), edge states are observed in Weyl semimetals. In particular,
  a pair of edge states meets along a line connecting two Weyl points of opposite chiralities, which is called Fermi arc \cite{Ferarc, Ferarc2}. Weyl semimetals are known to exhibit novel transport properties, such as negative magnetoresistance \cite{NMR, NMR2, NMR3}, anomalous Hall effect \cite{AHE, AHE1, AHE2, AHE3}, and chiral magnetic effect \cite{AHE3, CME1, CME2}. In 2015, a new type of Weyl semimetal phases called type-II Weyl semimetals was discovered \cite{wyel2}. In type-II Weyl semimetals, the energy dispersion near the Weyl points forms a tilted cone. As a result, the quasiparticle near these Weyl points behaves as a new type of quasiparticles which do not respect Lorentz invariance and thus have never been encountered in high energy physics. Moreover, type-II Weyl semimetals possess novel transport properties different from normal (type-I) Weyl semimetals. For example, in type-II Weyl semimetals, chiral anomaly exists only if the direction of the magnetic field is within the tiled cone \cite{wyel2} and the anomalous Hall effect depends on the tilt parameters \cite{wyel22}.

Despite the increasing efforts to realize these topological phases, engineering a controllable topological phase is quite challenging. One proposal to attain a controllable topological phase is to introduce a driving field (time periodic term) into a system. By using Floquet theory \cite{f01,f02,Floquet1, Floquet2}, it can be shown that such a driving field can modify the topology of the system's band structure. This method has been used to generate several topological phases such as Floquet topological insulators \cite{FTI, FTI2} and Floquet Weyl semimetals \cite{Fweyl}. Our recent studies have also shown how a variety of novel topological phases emerge in a periodically driven system \cite{Radit}. Note however, when the coupling with the driving field is sufficiently strong and the field itself is weak, then it becomes necessary to treat the driving field quantum mechanically as a collection of photons. On the one hand,
the total Hamiltonian including the photons has a larger dimension; on the other hand, it becomes time independent and our intuition about static systems can be useful again. This can sometimes offer an advantage over Floquet descriptions in the classical driving field case. As a result, several works have also been done on the constructions of nontrivial topological phases induced by a quantized field \cite{q1, q2}.

In this paper, we show another example of topological phase engineering via interaction with a driving field. Our starting static system is the generalized Harper model, i.e., Harper model with an off-diagonal modulation. This effectively one dimensional (1D) model has been known to simulate a Weyl semimetal phase with the help of its two periodic parameters which serve as artificial dimensions \cite{Radit, Sarma}. In our previous work \cite{Radit}, we have shown that adding a driving term in a form of a series of Dirac delta kicks leads to the emergence of new Weyl points. Here, we consider a more realistic driving term of the form $\propto \cos(\Omega t)$, with $\Omega$ being its frequency, to replace the kicking term in our previous model. As a result, our model is now more accessible experimentally. In addition, the simplicity of the model allows us to treat the driving term quantum mechanically and consider the full quantum picture of the system, which can then be compared with the semiclassical picture, i.e., by treating the particle quantum mechanically and the driving term classically. We find that when the driving term is treated classically, only type-I Weyl points are found. However, by treating the driving term quantum mechanically, some of the type-I Weyl points may turn into type-II Weyl points. Moreover, by tuning the coupling strength, we can control the tilt of the Weyl points and the energy difference between two Weyl points. This makes it possible to generate a pair of mixed Weyl points, with one belonging to type-I and the other belonging to type-II.

This paper is organized as follows. In Sec.~\ref{model}, we introduce the details of the model studied in this paper and set up some notation. In Sec.~\ref{type1}, we focus on the semiclassical case when the driving field is treated classically. We elucidate from both numerical and analytical perspectives how new type-I Weyl points emerge when the coupling strength is increased, and discuss its implications on the formation of edge states and quantization of adiabatic pump. In Sec.~\ref{compares}, we briefly explain the comparison with the static version of the model. In Sec.~\ref{type2}, we focus on the fully quantum version when the driving field is treated quantum mechanically. We show that the Weyl points are formed at the same locations as those in the semiclassical case. However, some of these Weyl points are now tilted and the energy at which they emerge is shifted by an amount which depends on the coupling strength. In Sec.~\ref{exper}, we briefly propose some possible experimental realizations. In Sec.~\ref{detect}, we examine a way to distinguish type-II Weyl points from type-I Weyl points in our system based on the formation of Landau levels when a synthetic magnetic field is applied \cite{LL}. In Sec.~\ref{conc}, we summarize our results and discuss possible further studies.

\section{The model}
\label{model}

In this paper, we focus on the following Hamiltonian,

\begin{equation}
H(t) = \sum_{n=1}^{N-1} \left\lbrace\left[J+(-1)^n\lambda \cos(\phi_y)\right]|n\rangle \langle n+1 | +h.c.\right\rbrace + \sum_{n=1}^N (-1)^n \left[V_1+V_2\cos(\Omega t)\right] \cos(\phi_z) |n\rangle \langle n |\;,
\label{Ham}
\end{equation}

\n where $n$ is the lattice site index, $N$ is the total number of lattice sites, $J$ and $\lambda$ are parameters related to the hopping strength, $V_1$ is the onsite potential, $V_2$ represents the coupling with the harmonic driving field, and $\Omega=\frac{2\pi}{T}$ with $T$ being the period of the driving field. The parameters $\phi_y$ and $\phi_z$ can take any value in $(-\pi,\pi]$, so that they can be regarded as the quasimomenta along two artificial dimensions { \cite{note}}. As a result, although Eq.~({\ref{Ham}}) is physically a 1D model, it can be used to simulate 3D topological phases. For example, if $V_2=0$, Eq.~(\ref{Ham}) reduces to the off-diagonal Harper model (ODHM), which has been shown to exhibit a topological Weyl semimetal phase \cite{Sarma, Radit}. For nonzero $V_2$, the system is effectively coupled to a periodic driving field, and thus its topological properties are expected to change depending on the values of $V_2$. We shall refer to this system as the continuously driven off-diagonal Harper model (CDODHM), which is a modification of the off-diagonal kicked Harper model (ODKHM) considered in our previous work \cite{Radit}.

Under periodic boundary conditions (PBC), Eq.~(\ref{Ham}) is invariant under translation by two lattice sites. Therefore, Eq.~(\ref{Ham}) can be expressed in terms of the quasimomentum $k$ by using Fourier transform as

\begin{equation}
H(t) = \sum_k \mathcal{H}_k (t) \otimes |k\rangle \langle k |\;,
\end{equation}

\n where $|k\rangle$ is a basis state representing the quasimomentum $k$, and $\mathcal{H}_k$ is the momentum space Hamiltonian given by

\begin{eqnarray}
\mathcal{H}_k (t) &=& 2J\cos(k)\sigma_x +2\lambda\cos(\phi_y)\sin(k) \sigma_y +\left[V_1+V_2\cos(\Omega t)\right] \cos(\phi_z) \sigma_z \nonumber \\
&=& \mathcal{H}_{k,0}+V_2\cos(\Omega t) \cos(\phi_z) \sigma_z \;,
\label{Ham2}
\end{eqnarray}

\n with $\sigma_x$, $\sigma_y$, and $\sigma_z$ are Pauli matrices representing the sublattice degrees of freedom.

\section{Classical driving field}

\subsection{Emergence of type-I Weyl points}
\label{type1}

\begin{figure}
\begin{center}
\includegraphics[scale=0.3]{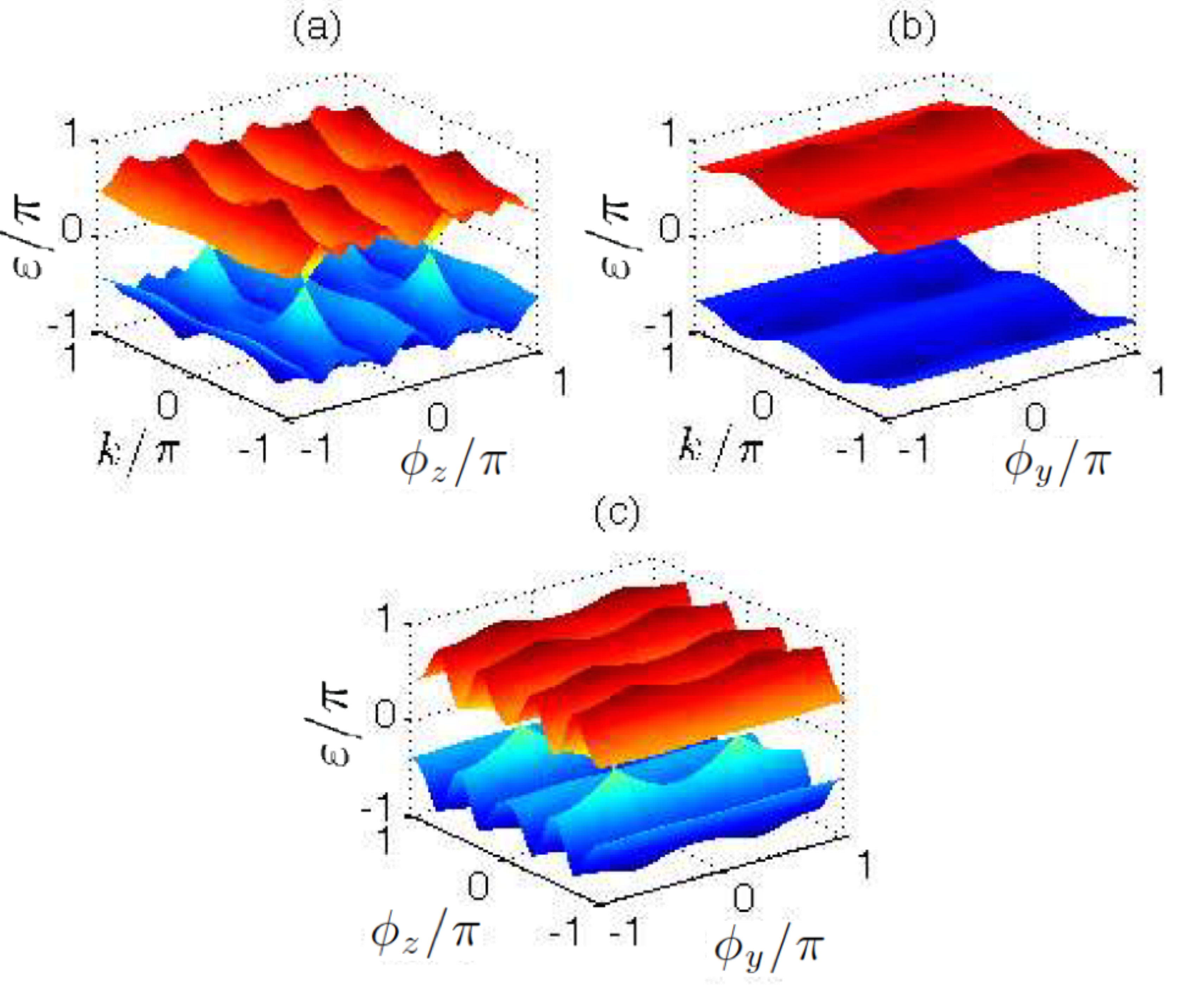}
\end{center}
\caption{(color online). A typical quasienergy spectrum of the CDODHM under PBC at a fixed (a) $\phi_y=\frac{\pi}{2}$, (b) $\phi_z=\arccos\left(\frac{\pi}{5}\right)$, and (c) $k=\frac{\pi}{2}$. Parameters used are $J=1$, $\lambda=0.5$, $V_1=5$, and $V_2=12$. Here and in all other figures, plotted quantities are in dimensionless units.}
\label{floquetpbc}
\end{figure}

Since the Hamiltonian described by Eq.~(\ref{Ham}) is time periodic, its properties can be captured by diagonalizing its corresponding Floquet operator $\left(U\right)$, which is defined as a one period time evolution operator. In particular, under PBC, by diagonalizing the momentum space Floquet operator $\left(\mathcal{U}_k\right)$ as a function of $k$, $\phi_y$, $\phi_z$ over the whole 3D Brillouin zone, i.e., the region $(-\pi,\pi]\times (-\pi,\pi]\times (-\pi,\pi]$ (with the lattice constant set to 1 for simplicity), we can obtain its Floquet band (quasienergy band). Fig.~\ref{floquetpbc} shows a typical quasienergy spectrum of the CDODHM in the unit where $T=\hbar=1$ and the parameters $J$, $\lambda$, $V_1$, and $V_2$ are dimensionless. Here, the quasienergy ($\varepsilon$) is defined as the phase of the eigenvalue of the Floquet operator, i.e., $\mathcal{U}_k |\psi\rangle = \mathrm{exp}\left(\mathrm{i} \varepsilon \right)|\psi \rangle$. By construction, $\varepsilon$ is only defined up to a modulus of $2\pi$, and thus $\varepsilon=-\pi$ and $\varepsilon=\pi$ are identical. Therefore, unlike the ODHM, which only exhibits band touching points at energy $0$, in the CDODHM, it is possible for the two bands to touch at both quasienergy $0$ and $\pi$, which is evident from Fig.~\ref{floquetpbc}.

Near these band touching points, time dependent perturbation theory can be applied to obtain an approximate analytical expression of the momentum space Floquet operator. By leaving any technical details in Appendix \ref{app1}, it is found that the momentum space Floquet operator around a band touching point at $(k,\phi_y,\phi_z)=\left(\frac{\pi}{2},\frac{\pi}{2}, \phi_l\right)$, with $\phi_l=\arccos\left(\frac{l\pi}{V_1}\right)$, is given by

\begin{equation}
\mathcal{U}(k_x, k_y, k_z) = \exp\left\lbrace-\mathrm{i}\left\lbrace l\pi-\left[ V_1 k_z\sin(\phi_l)\sigma_z+2Jk_x J_l(l c)\sigma_x +2\lambda k_y J_l(l c)\sigma_y\right]\right\rbrace\right\rbrace \;,
\label{Uw}
\end{equation}

\n where $k_x\equiv k-\frac{\pi}{2}$, $k_y\equiv\phi_y-\frac{\pi}{2}$, $k_z\equiv\phi_z-\phi_l$, $c=\frac{V_2}{V_1}$, and $J_l$ is the Bessel function of the first kind. By comparing Eq.~(\ref{Uw}) with the general form $\mathcal{U}=\exp\left[-\mathrm{i} \mathcal{H}_\mathrm{eff}\right]$ of the momentum space Floquet operator, with $\mathcal{H}_\mathrm{eff}$ be the momentum space effective Hamiltonian, it is found that

\begin{equation}
\mathcal{H}_\mathrm{eff} = l\pi-\left[ V_1 k_z\sin(\phi_l)\sigma_z+2Jk_x J_l(l c)\sigma_x +2\lambda k_y J_l(l c)\sigma_y\right] \;.
\label{effH}
\end{equation}

\n Eq.~(\ref{effH}) is in the form of a Weyl Hamiltonian with chirality $\chi = -\mathrm{sgn}\left[V_1J\lambda \sin(\phi_l)\right]$ \cite{Hosur} and quasienergy

\begin{equation}
\varepsilon = \left\lbrace\begin{array}{cc} \pm\left[\pi-\sqrt{V_1^2k_z^2\sin^2(\phi_l)+4J^2k_x^2 J_l^2(lc)+4\lambda^2k_y^2 J_l^2(lc)}\right] & \mathrm{if}\,\, l \,\, \mathrm{is}\,\, \mathrm{odd} \\ \pm\sqrt{V_1^2k_z^2\sin^2(\phi_l)+4J^2k_x^2 J_l^2(lc)+4\lambda^2k_y^2 J_l^2(lc)} & \mathrm{if}\,\, l \,\, \mathrm{is}\,\, \mathrm{even}\end{array}\right. \;.
\label{effe}
\end{equation}

\n In particular, because of the absence of any tilting term \cite{wyel2} in Eq.~(\ref{effH}), it describes a type-I Weyl Hamiltonian. Consequently, the band touching point at $(k,\phi_y,\phi_z)=\left(\frac{\pi}{2},\frac{\pi}{2}, \phi_l\right)$ corresponds to a type-I Weyl point.

\begin{figure}
\begin{center}
\includegraphics[scale=0.3]{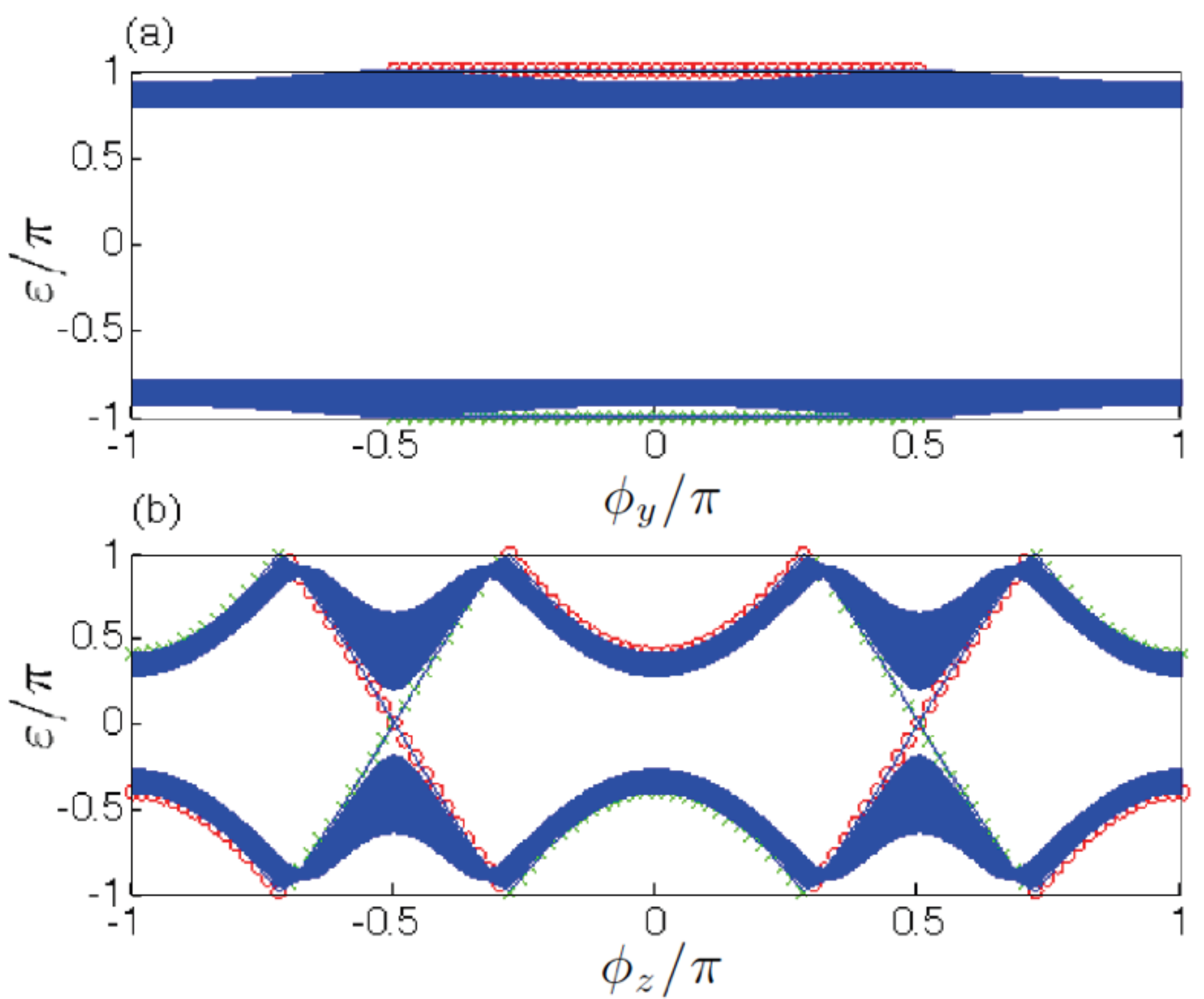}
\end{center}
\caption{(color online). A typical quasienergy spectrum of the CDODHM under OBC at a fixed (a) $\phi_z=\arccos\left(\frac{\pi}{5}\right)$ and (b) $\phi_y=0.3\pi$. Parameters used are $J=1$, $\lambda=0.5$, $V_1=5$, $V_2=2$, and $N=100$. Red circles and green crosses denote edge states localized around the right and left end, respectively.}
\label{floquetobc}
\end{figure}

In order to verify their topological signature, Fig.~\ref{floquetobc} shows the quasienergy spectrum of the Floquet operator associated with Eq.~(\ref{Ham}) under OBC, i.e., by taking a finite $N=100$. Fig.~\ref{floquetobc}a shows that two dispersionless edge states (marked by red circles and green crosses) emerge at quasienergy $\pi$ connecting two Weyl points with opposite chiralities when viewed at a fixed $\phi_z$. These edge states are analogues to Fermi arcs in static Weyl semimetal systems \cite{Ferarc}, and they arise as a consequence of the topology of the Floquet Su-Schrieffer-Heeger (SSH) model \cite{Radit}. When viewed at a constant $|\phi_y|<\frac{\pi}{2}$, as shown in Fig.~\ref{floquetobc}b, two edge states are shown to traverse the gap between the two Floquet bands and meet at the projection of the Weyl points onto the plane of constant $\phi_y$. These edge states emerge due to the topology of two mirror copies of Floquet Chern insulators \cite{Sarma}, and disappear when $|\phi_y|>\frac{\pi}{2}$, due to the topological transition from Floquet Chern to normal insulators. Floquet Fermi arcs observed in Fig.~\ref{floquetobc}a are formed by joining these meeting points starting from the plane $\phi_y=-\frac{\pi}{2}$ to the plane $\phi_y=\frac{\pi}{2}$, i.e., the locations of two Weyl points with opposite chiralities. The 3D nature of the CDODHM can therefore be constructed by stacking a series of Floquet Chern insulators sandwiched by normal insulators. The Weyl points emerge at the interface separating the Floquet Chern and normal insulators.

The topological charge (chirality) of the Weyl points can also be manifested in terms of the quantization of adiabatic transport. According to our previous work \cite{Radit}, by preparing a certain initial state and driving it adiabatically along a closed loop in the parameter space ($\phi_y$ and $\phi_z$), the change in position expectation value after one full cycle is given by

\begin{equation}
\Delta \langle X\rangle = a \chi_\mathrm{enc}\;,
\label{exp}
\end{equation}

\n where $a$ is the effective lattice constant, which is equal to $2$ in this case since one unit cell consists of two lattice sites, and $\chi_\mathrm{enc}$ is the total chirality of the Weyl points enclosed by the loop. By following the same procedure in Ref.~\cite{Radit}, we prepare the following initial state,

\begin{equation}
| \Psi(t=0) \rangle = \frac{1}{2\pi}\int_{-\pi}^\pi |\psi_{-}(k,\phi_y(0),\phi_z(0)) \rangle dk \;,
\label{is}
\end{equation}

\begin{figure}
\begin{center}
\includegraphics[scale=0.4]{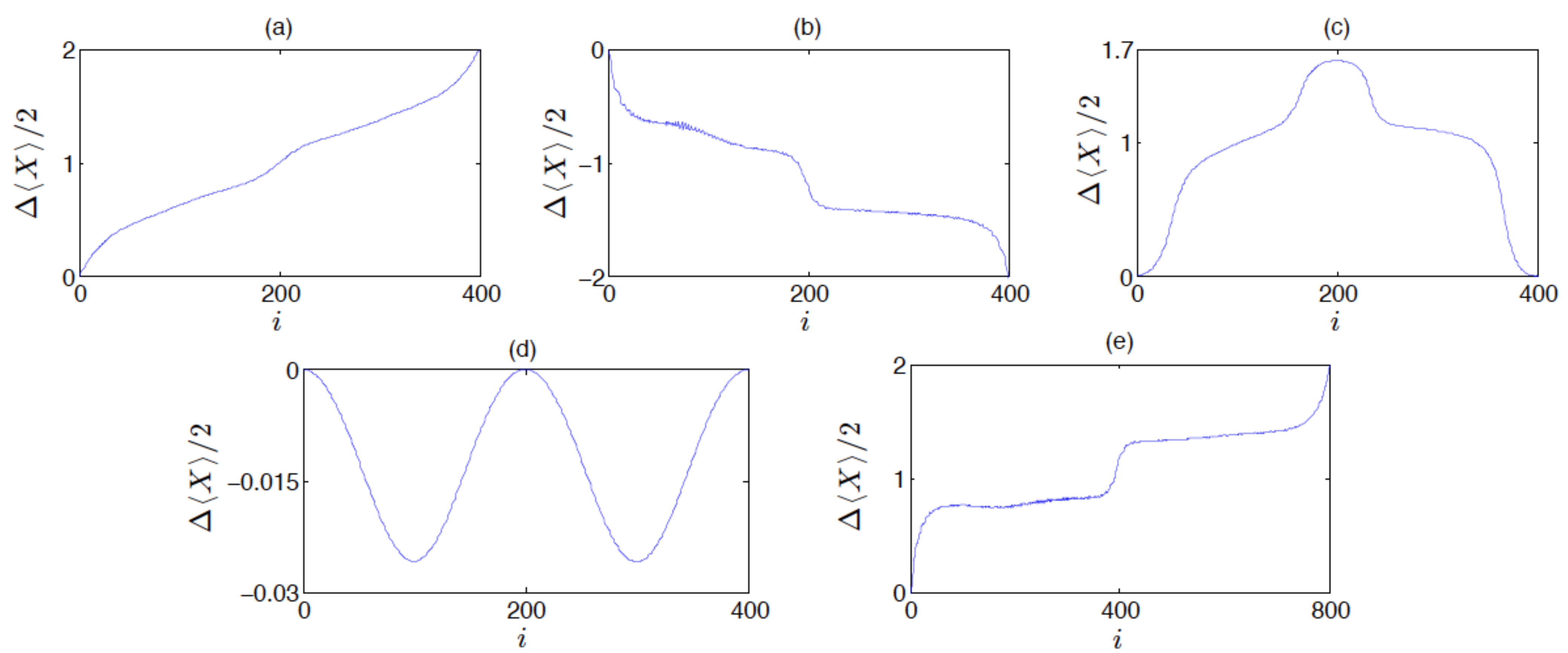}
\end{center}
\caption{The change in position expectation value as a function of $i$ when Eq.~(\ref{is}) is driven adiabatically along various loops in the parameter space. The loops are chosen to enclose Weyl points at (a) $(k,\phi_y,\phi_z)=\left(\pm\frac{\pi}{2},\frac{\pi}{2},\frac{\pi}{2}\right)$, (b) $(k,\phi_y,\phi_z)=\left(\pm\frac{\pi}{2},\frac{\pi}{2},\phi_1\right)$, (c) $(k,\phi_y,\phi_z)=\left(\pm\frac{\pi}{2},\frac{\pi}{2},\frac{\pi}{2}\right)$ and $(k,\phi_y,\phi_z)=\left(\pm\frac{\pi}{2},\frac{\pi}{2},\phi_1\right)$, (d) no Weyl point, and (e) new Weyl points at $(k,\phi_y,\phi_z)=\left(\pm\frac{\pi}{2},\frac{\pi}{2},\phi_2\right)$ emerging when $V_1>2\pi$. The parameters chosen are $J=1$, $\lambda=0.5$, (a) and (d) $V_1=\frac{V_2}{2}=2$, (b) and (c) $V_1=\frac{V_2}{2}=5$, (e) $V_1=\frac{V_2}{2}=7$. In (a)$-$(d), $N=400$, whereas in (e), $N=800$.}
\label{pump1}
\end{figure}

\n where $|\psi_-(k,\phi_y,\phi_z)$ is the Floquet eigenstate associated with the lower band in Fig.~\ref{pump1}, and $\phi_y$ and $\phi_z$ are tuned adiabatically according to $\phi_y=\phi_{y,0}+r\cos[\theta(t)+\Phi]$ and $\phi_z=\phi_{z,0}+r\sin[\theta(t)+\Phi]$, with $\Phi$ be a constant phase and $\theta(t)=\frac{2\pi i}{M}$ for $i-1<t\leq i$ and $0<i\leq M$. Adiabatic condition is reached by setting $M$ to be very large. Fig.~\ref{pump1} shows the change in position expectation value of Eq.~(\ref{is}) after it is driven along various closed loops in parameter space. It is evident from the figure that Eq.~(\ref{exp}) is satisfied. For instance, when the loop is chosen to enclose two Weyl points with the same chiralities, i.e., Fig.~\ref{pump1}a, \ref{pump1}b, and \ref{pump1}e, $\frac{\Delta \langle X\rangle }{2}=\pm 2$ after one full cycle, whereas if it encloses Weyl points with opposite chiralities or no Weyl point, i.e., Fig.~\ref{pump1}c and \ref{pump1}d, $\frac{\Delta \langle X\rangle }{2}=0$ after one full cycle.

\subsection{Comparison with the ODHM}
\label{compares}

According to our findings in Sec.~\ref{type1}, the CDODHM is able to host as many type-I Weyl points as possible by simply increasing the parameter $V_1$. As $V_1$ increases, there are more integers satisfying $l\pi \leq V_1$, and hence more type-I Weyl points emerge. On the other hand, in the ODHM, i.e., $V_2=0$ case, no matter what the values of the parameters $J$, $\lambda$, and $V_1$ are, there are only 8 type-I Weyl points touching at energy $0$, corresponding to $(k,\phi_y,\phi_z)=\left(\pm\frac{\pi}{2}, \pm\frac{\pi}{2}, \pm\frac{\pi}{2}\right)$. This can be understood as follows. If $l\neq 0$, then $J_l(lc)=J_l(0)=0$, which implies that terms proportional to Pauli matrices $\sigma_x$ and $\sigma_y$ in Eq.~(\ref{Uw}) are missing. As a result, Eq.~(\ref{effH}) no longer describes a Weyl Hamiltonian, and the band touching point at $(k,\phi_y,\phi_z)=\left(\pm\frac{\pi}{2},\pm\frac{\pi}{2}, \phi_l\right)$ for $l\neq 0$ is not a Weyl point. If however $l=0$, i.e., $\phi_l=\phi_0=\pm \frac{\pi}{2}$, then $J_l(lc)=J_0(0)=1$, and the terms proportional to Pauli matrices $\sigma_x$ and $\sigma_y$ in Eq.~(\ref{Uw}) remain nonzero. Consequently, Eq.~(\ref{effH}) still describes a type-I Weyl Hamiltonian, and the band touching point at $(k,\phi_y,\phi_z)=\left(\pm\frac{\pi}{2},\pm\frac{\pi}{2}, \pm \frac{\pi}{2}\right)$ corresponds to a type-I Weyl point.

The emergence of the additional type-I Weyl points in the CDODHM can be understood as follows. First, we separate the time dependent and independent part of Eq.~(\ref{Ham2}). The time independent part is simply the ODHM momentum space Hamiltonian, whereas the time dependent part can be understood as its interaction with the driving field, which can in general induce transition between the two energy bands of the ODHM, and hence modify its band structure. When $V_1\geq l\pi$, there exists a point in the Brillouin zone at which the energy difference between the two bands of the ODHM is equal to $2l\pi$. In the unit we choose, this energy difference also represents the transition frequency between the two energy levels, which is on resonance with the frequency of the driving field $\Omega=2\pi$. As a result, the two energy levels will be dynamically connected with each other, yielding a type-I Weyl point in the quasienergy spectrum.

\section{Quantum driving field}

\subsection{Quantized model}

Quantum mechanically, the Hamiltonian of the driving field takes the form of the harmonic oscillator Hamiltonian, which can be written as
\begin{equation}
H_{\rm field}= \Omega a^\dagger a\;,
\label{field}
\end{equation}

\n where $a$ ($a^\dagger$) is the photon destruction (creation) operator, and the zero point energy $\frac{1}{2} \Omega$ has been suppressed since it will not contribute to our present analysis. In the Heisenberg picture, the time dependence of $a$ and $a^\dagger$ can be found by solving the following equation of motion,
\begin{eqnarray}
\frac{da}{dt} &=&-\frac{ \left[H_\mathrm{field}, a\right]}{\mathrm{i}} \nonumber \\
&=& -\mathrm{i} \Omega a\; .
\label{em}
\end{eqnarray}

\n It can be immediately verified from Eq.~(\ref{em}) that $a(t)=a(0)\exp\left(-\mathrm{i} \Omega t \right)$ and $a^\dagger(t)=a^\dagger(0)\exp\left(\mathrm{i} \Omega t \right)$. By including the quantized driving field as part of our system, the total Hamiltonian can be written as
\begin{equation}
H_\mathrm{tot} = I_p \otimes H_\mathrm{ODHM}+H_\mathrm{field}\otimes I_\mathrm{ODHM} + H_\mathrm{int} \;,
\label{Htot}
\end{equation}

\n where $H_\mathrm{ODHM}$ is the ODHM Hamiltonian (the time independent part of Eq.~(\ref{Ham})), $I_p$ and $I_\mathrm{ODHM}$ are the identity operator in the photon and the ODHM space respectively, and $H_\mathrm{int}$ is the interaction Hamiltonian describing the coupling between the ODHM and the driving field. The form of $H_\mathrm{int}$ can be obtained from the time dependent part of Eq.~(\ref{Ham}). By writing $\cos\left(\Omega t\right) =\frac{1}{2}\left[\exp\left(-\mathrm{i} \Omega t \right)+\exp\left(\mathrm{i} \Omega t \right)\right]$ in Eq.~(\ref{Ham}), we can identify $\exp\left(-\mathrm{i} \Omega t \right)$ and $\exp\left(\mathrm{i} \Omega t \right)$ terms as $a(t)$ and $a^\dagger(t)$ respectively. The time dependence of $a$ and $a^\dagger$ can be transferred to the corresponding basis states in the photon space (by changing from the Heisenberg to the Schrodinger picture) \cite{Loudon}, so that Eq.~(\ref{Htot}) is time independent, with $H_\mathrm{int}$ given by
\begin{equation}
H_\mathrm{int} = \sum_n^N (-1)^n \frac{V_2\cos(\phi_z)}{2} \left(a+a^\dagger \right) \otimes |n\rangle \langle n | \; .
\label{Hint}
\end{equation}

Under PBC, the momentum space Hamiltonian associated with Eq.~(\ref{Htot}) is given by
\begin{equation}
\mathcal{H}_{\rm tot} = I_p \otimes \mathcal{H}_\mathrm{k,0}+H_\mathrm{field}\otimes I_\mathrm{2} + \mathcal{H}_\mathrm{int} \;,
\label{Htotk}
\end{equation}

\n where $\mathcal{H}_{k,0}$ is given by Eq.~(\ref{Ham2}), $I_2$ is a $2\times 2$ identity matrix, and

\begin{equation}
\mathcal{H}_\mathrm{int} = \frac{V_2\cos(\phi_z)}{2} \left(a+a^\dagger \right) \otimes \sigma_z \;.
\end{equation}

\begin{figure}
\begin{center}
\includegraphics[scale=0.45]{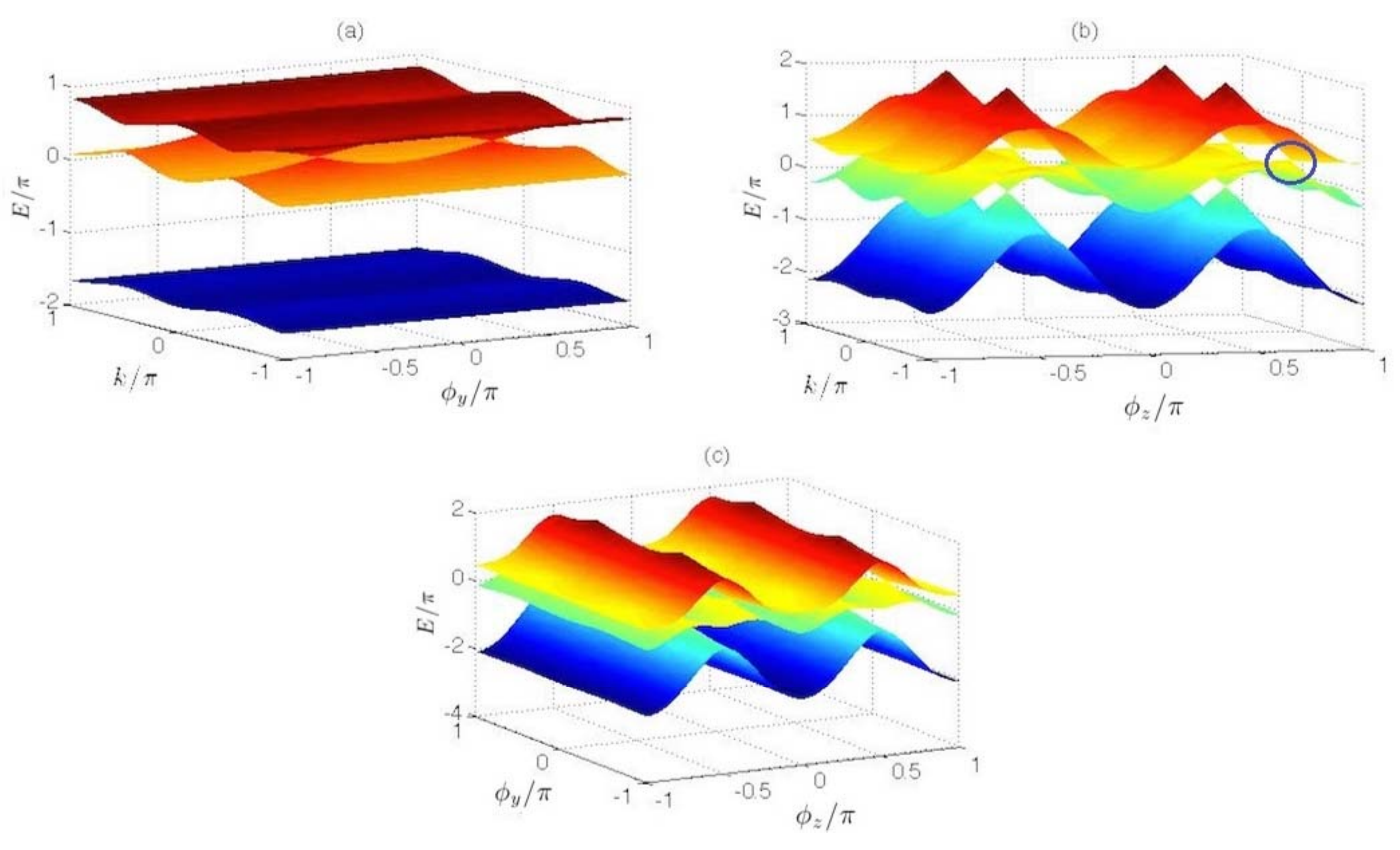}
\end{center}
\caption{(color online). A typical energy spectrum of the quantized CDODHM under PBC for the first three bands at a fixed (a) $\phi_z=0.2\pi$, (b) $\phi_y=\frac{\pi}{2}$, and (c) $k=\frac{\pi}{2}$. Parameters used are $J=1$, $\lambda=0.5$, $V_1=\frac{\pi}{\cos(0.2\pi)}$, $V_2=8$, and the photon number is truncated at $N_p=50$. The blue circle in (b) highlights the tilted cone in the energy spectrum.}
\label{quanpbc}
\end{figure}

\begin{figure}
\begin{center}
\includegraphics[scale=0.3]{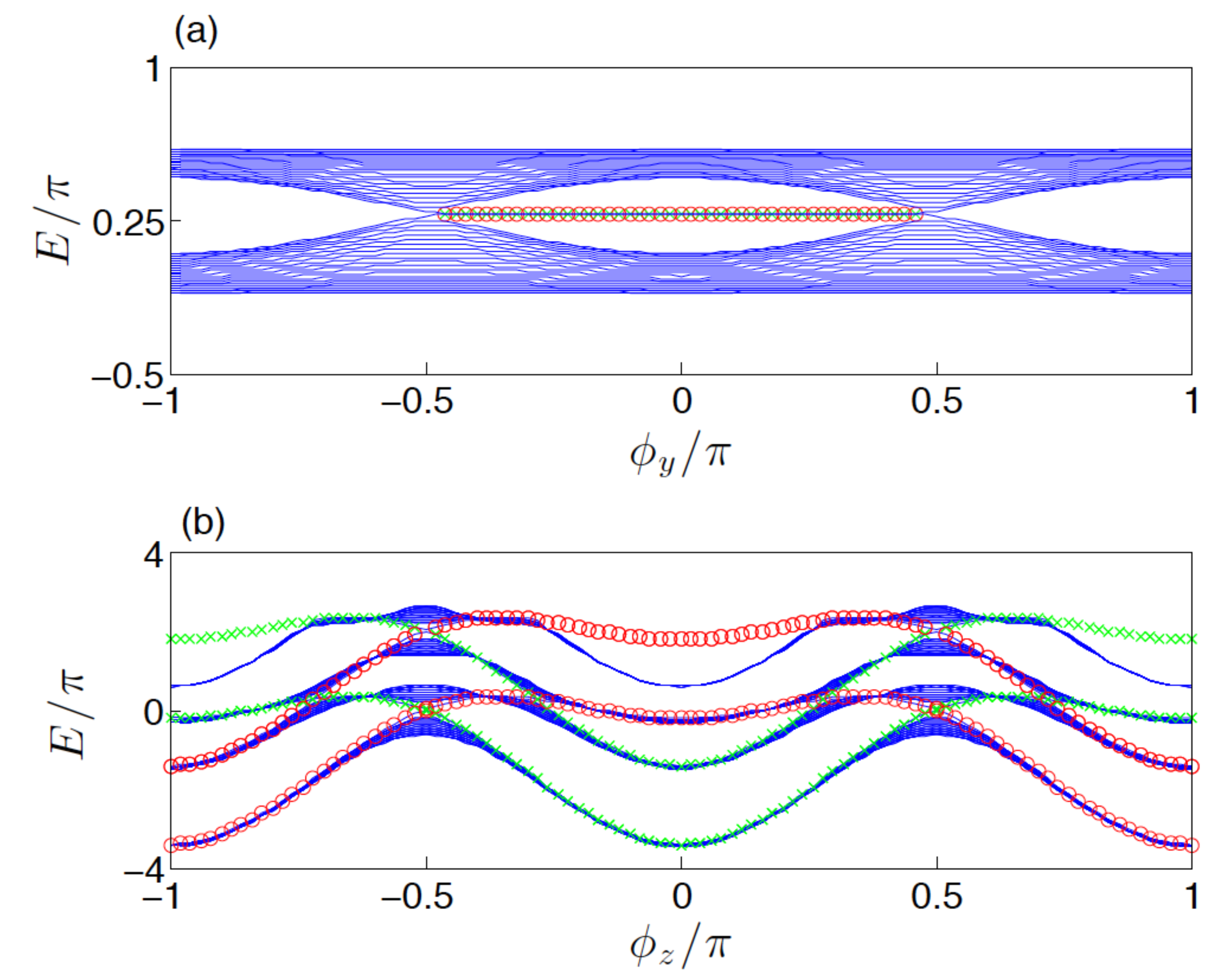}
\end{center}
\caption{(color online). A typical energy spectrum of the quantized CDODHM under OBC at a fixed (a) $\phi_z=\arccos\left(\frac{\pi}{5}\right)$ and (b) $\phi_y=0.3\pi$. Parameters used are $J=1$, $\lambda=0.5$, $V_1=5$, $V_2=12$, $N=50$, and the photon number is truncated at $N_p=10$. (a) shows the second and third energy bands, while (b) shows the first four energy bands. Red circles and green crosses denote edge states localized around the right and left end, respectively. }
\label{quanobc}
\end{figure}

\n Fig.~\ref{quanpbc} and Fig.~\ref{quanobc} show a typical energy band structure of the model under PBC and OBC, obtained by diagonalizing Eq.~(\ref{Htotk}) and Eq.~(\ref{Htot}), respectively. It is observed from Fig.~\ref{quanpbc} that in addition to the Weyl points at $(k,\phi_y,\phi_z)=(\pm\frac{\pi}{2},\pm\frac{\pi}{2},\pm\frac{\pi}{2})$, new Weyl points emerge at some other points. Fermi arc surface states connecting each pair of these new Weyl points, similar to what we observed in Sec.~\ref{type1}, are also evident from Fig.~\ref{quanobc}a, which confirms their topological nature. Near these new Weyl points, the energy dispersion forms a tilted cone (blue circle in Fig.~\ref{quanpbc}b), suggesting that they might be categorized as type-II Weyl points. In Sec.~\ref{type2}, we are going to show analytically that these type-II Weyl points emerge at the same points as the additional type-I Weyl points were the driving field treated classically, as elucidated in Sec.~\ref{type1}. This result suggests that in the quantum limit, some type-I Weyl points will turn into type-II Weyl points.

\subsection{Emergence of type-II Weyl points}
\label{type2}

Although Eq.~(\ref{Htotk}) is time independent, it now has a larger dimension since it includes the photon space. By introducing the quadrature operators $X$ and $P$ satisfying the commutation relation $[X,P]=\mathrm{i}$ and are related to $a$ and $a^\dagger$ by
\begin{eqnarray}
a &=& \frac{1}{\sqrt{2}}\left(X+\mathrm{i} P\right) \;,\\
a^\dagger &=& \frac{1}{\sqrt{2}}\left(X-\mathrm{i} P\right) \;,
\end{eqnarray}

\n Eq.~(\ref{Htotk}) becomes (at $k=\phi_y=\frac{\pi}{2}$)

\begin{equation}
\mathcal{H}_{\mathrm{tot},\pm}\left(\frac{\pi}{2},\frac{\pi}{2},\phi_z\right)=2\pi \left\lbrace \frac{P^2}{2}+\frac{1}{2}\left[X\pm \frac{V_2\cos(\phi_z)}{2\sqrt{2}\pi}\right]^2-\frac{1}{2}\right\rbrace \pm V_1\cos(\phi_z) -\frac{V_2^2\cos^2(\phi_z)}{8\pi}\;.
\label{osc}
\end{equation}

\n Eq.~(\ref{osc}) is simply the harmonic oscillator Hamiltonian with shifted ``position" expectation value. Near $(k,\phi_y,\phi_z)=\left(\frac{\pi}{2},\frac{\pi}{2}, \phi_1 \right)$, with $\phi_1=\arccos\left(\frac{\pi}{V_1}\right)$, it is shown in Appendix \ref{app2} that the energy dispersion is given by

\begin{eqnarray}
E_{n,\pm} &=& \pi (2n-1)-\frac{\pi V_2^2}{8V_1^2} +\frac{V_2^2}{4V_1}k_z \sin(\phi_1) \nonumber \\
&& \pm \sqrt{V_1^2 \sin^2(\phi_1) k_z^2+4J^2J_1\left(\frac{\sqrt{n}V_2}{V_1}\right)^2 k_x^2+4\lambda^2 J_1\left(\frac{\sqrt{n}V_2}{V_1}\right)^2 k_y^2} \;,
\label{weyl2}
\end{eqnarray}

\n where, similar to our previous notation,  $k_x=k-\frac{\pi}{2}$, $k_y=\phi_y-\frac{\pi}{2}$, and $k_z=\phi_z-\phi_1$. Furthermore, Eq.~(\ref{Htotk}) will be block diagonal in the basis spanned by the eigenstates associated with Eq.~(\ref{unperbE}) in Appendix \ref{app2}, where each subblock consists of $2\times 2$ matrix which can be written in the following form,

\begin{equation}
[\mathcal{H}_q]_n = \pi (2n-1)-\frac{\pi V_2^2}{8V_1^2} +\frac{V_2^2}{4V_1}k_z \sin(\phi_1)-V_1 \sin(\phi_1) k_z\tau_z -\left(2Jk_x \tau_x + 2\lambda k_y \tau_y\right) J_1\left(\frac{\sqrt{n}V_2}{V_1}\right)\;,
\label{weyl}
\end{equation}

\n where $\tau_x$, $\tau_y$, and $\tau_z$ take the form of Pauli matrices. Eq.~(\ref{weyl}) is in the form of a Weyl Hamiltonian, which resembles a similarity with Eq.~(\ref{effH}) found in Sec.~\ref{type1}, apart from the extra tilting term $\frac{V_2^2}{4V_1}k_z \sin(\phi_1)$ and the energy shift $-\frac{\pi V_2^2}{8V_1^2}$. These extra terms in turn lead to novel phenomena which are not captured if the driving field is treated classically. First, because of the tilting term, it is possible for the Dirac cone around the Weyl point described by Eq.~(\ref{weyl}) to tip over at a sufficiently large matter-field coupling $V_2$, so that it is categorized into type-II Weyl points. According to the classification in Ref.~\cite{wyel2}, this Weyl point is a type-II Weyl point if $V_2>2V_1$. Second, the energy shifting term will shift the energy at which the Weyl point is formed, so that it is not an integer multiple of $\pi$.

These two phenomena are the main results of this paper, which have some fascinating implications. First, they show the difference between quantum and classical treatments of light, which is one of the main interests in the studies of quantum optics \cite{Loudon}. Second, since both the tilting and energy shifting terms are proportional $\propto V_2^2$, they can be easily controlled by simply tuning $V_2$. Moreover, we note that these two terms will not affect the Weyl points at $\left(k,\phi_y,\phi_z\right)=\left(\pm \frac{\pi}{2}, \pm \frac{\pi}{2}, \pm \frac{\pi}{2} \right)$, which can be easily verified by expanding Eq.~(\ref{osc}) up to first order near these points. By following the same procedure that leads to Eq.~(\ref{weyl}), it can be shown that both the second (the energy shifting) and the third (the tilting) terms are missing. As a result, these Weyl points always correspond to type-I Weyl points and are located at a fixed energy regardless of $V_2$. This implies that by tuning $V_2$, it is possible to generate a pair of mixed Weyl points, with one belonging to type-I while the other belonging to type-II, separated by a controllable energy difference. This might serve as a good starting point to study further the properties of such mixed Weyl semimetal systems. For example, by fixing $\phi_y$ and $\phi_z$ in between a pair of mixed Weyl points and applying a magnetic field, one could explore the possiblity of generating the chiral magnetic effect \cite{AHE3,CME1,CME2}, i.e., the presence of dissipationless current along the direction of the magnetic field, which is known to depend on the energy difference between two type-I Weyl points \cite{Hosur,AHE3}.

\begin{figure}
	\begin{center}
		\includegraphics[scale=0.35]{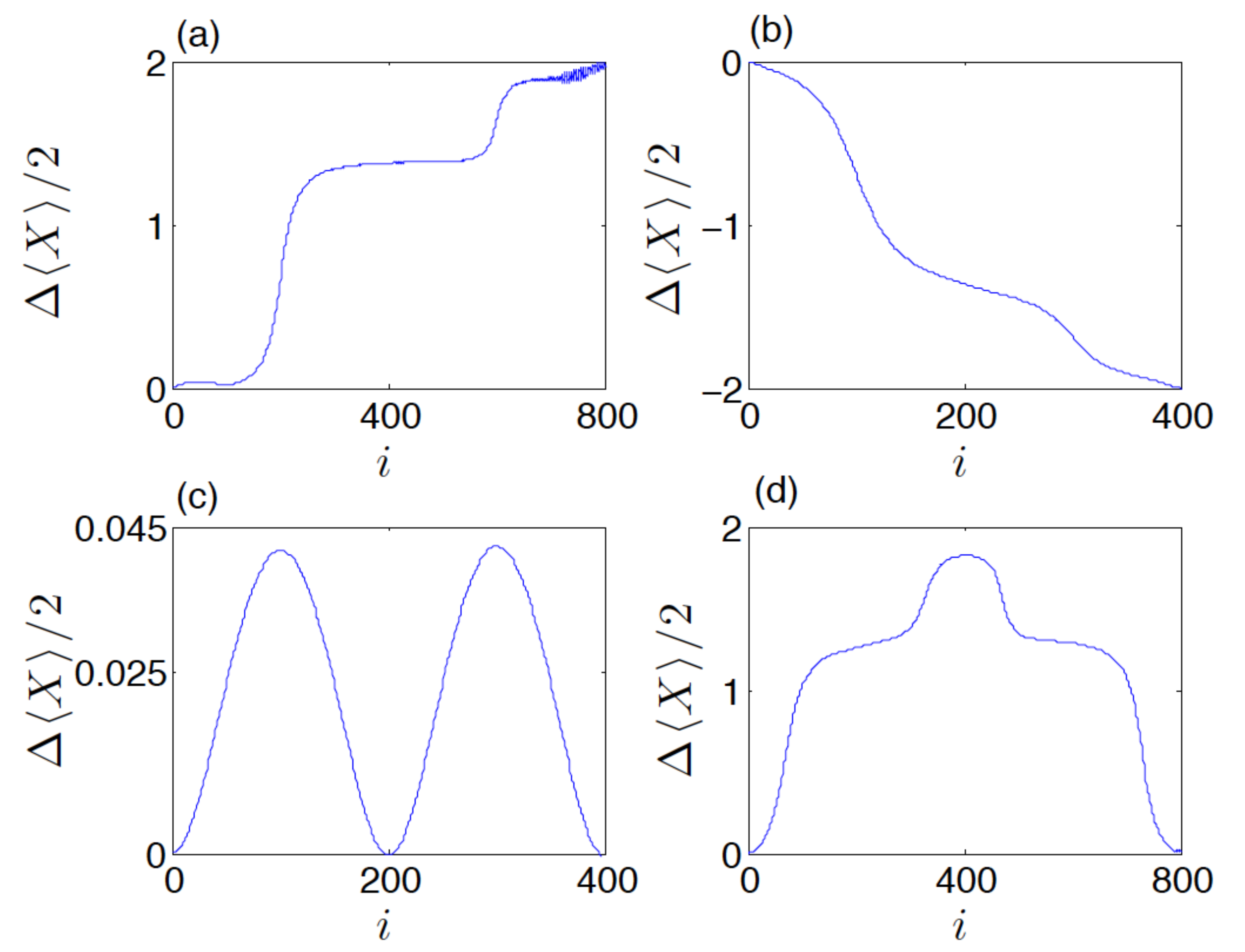}
	\end{center}
	\caption{The change in position expectation value as a function of $i$ when an initial state similar to Eq.~(\ref{is}) is driven adiabatically along various loops in the parameter space. The loops are chosen to enclose Weyl points at (a) $(k,\phi_y,\phi_z)=\left(\pm\frac{\pi}{2},\frac{\pi}{2},\arccos\left[\frac{\pi}{V_1}\right]\right)$, (b) $(k,\phi_y,\phi_z)=\left(\pm\frac{\pi}{2},\frac{\pi}{2},\frac{\pi}{2}\right)$, (c) no Weyl point, and (d) Weyl points at $(k,\phi_y,\phi_z)=\left(\pm\frac{\pi}{2},\frac{\pi}{2},\arccos\left[\frac{\pi}{V_1}\right]\right)$ and $(k,\phi_y,\phi_z)=\left(\pm\frac{\pi}{2},\frac{\pi}{2},\frac{\pi}{2}\right)$. The parameters chosen are $J=1$, $\lambda=0.5$, (a) and (d) $V_1=5$ and $V_2=12$, (b) and (c) $V_1=V_2=2$. In (a) and (d), $N=800$, whereas in (b) and (c), $N=400$.}
	\label{pump2}
\end{figure}

Despite the difference between the semiclassical and fully quantum results described above, they share some similarities in terms of the quantization of adiabatic transport. Fig.~\ref{pump2} shows the change in position of expectation value when an initial state similar to Eq.~(\ref{is}) is driven adiabatically along various closed loops by tuning $\phi_y$ and $\phi_z$ in the same manner as that elucidated in Sec.~\ref{type1}. Similar to what we observed in Sec.~\ref{type1}, the change in position expectation value after one full cycle still obeys Eq.~(\ref{exp}) regardless of the type of the Weyl points enclosed. This indicates clearly that a transition from type-I to type-II Weyl point will preserve its chirality. This makes sense since such a transition is induced by a term that doesn't depend on any of the Pauli matrices, and hence will not affect its chirality.

We end this section by presenting a comparison between Eq.~(\ref{weyl}) and Eq.~(\ref{effH}). By identifying $V_2$ in Eq.~(\ref{effH}) as $V_2\sqrt{n}$ in Eq.~(\ref{weyl}), it can be immediately shown that Eq.~(\ref{weyl}) will reduce to Eq.~(\ref{effH}) when $V_2\rightarrow 0$ while $n\rightarrow \infty$, such that $V_2\sqrt{n}$ remains finite. In this regime, Eq.~(\ref{weyl2}) will be periodic with a modulus of $2\pi$, which is the same as Eq.~(\ref{effe}). This explains why the extra tilting and energy shifting terms are not observed in the classical driving field case. Since these two terms are proportional to $V_2^2$, their effect will diminish as we move from the quantum to classical driving field regime. This observation can also be understood more physically as follows. In both quantum and classical field regime, the additional Weyl points emerge as a result of the resonance between the particle transition frequency and the frequency of the driving field. Since the interaction between the particle and a single photon depends on the parameter $\phi_z$, it is expected in general that the modification of the band structure near the resonant points (the additional Weyl points) also depends on $\phi_z$, resulting in the emergence of the tilting term in the full quantum field regime. Since the Weyl points at $(k,\phi_y,\phi_z)=\left(\pm\frac{\pi}{2}, \pm\frac{\pi}{2}, \pm\frac{\pi}{2}\right)$ are not resonant with the driving field, the interaction effect will be quite small, and the $\phi_z$ dependence effect of the interaction will not be visible near these Weyl points, which explains the absence of the tilting term even in the full quantum field regime. The energy shifting term in the full quantum field regime is a result of the change in the energy difference between the Weyl points at $(k,\phi_y,\phi_z)=\left(\pm\frac{\pi}{2}, \pm\frac{\pi}{2}, \pm\frac{\pi}{2}\right)$ and the resonant points before and after the driving field is introduced. Finally, in the classical field regime, the interaction between the particle and a single photon is very weak. Although there are infinitely many photons in the classical field case, both the tilting and energy shifting terms depend only on the interaction strength with a single photon even near the resonant points. Therefore, the most visible effect of the interaction with all the photons is to just dress the band structure near the Weyl points, which is uniform up to first order in $\phi_z$.

\section{Discussions}

\subsection{Possible experimental realizations}\label{exper}

There have already been several proposals to experimentally realize the Harper model in the framework of ultracold atom systems \cite{exp0,exp01} as well as optical waveguides \cite{exp1, exp2}. The semiclassical version of our model can be easily realized by slightly modifying some of these experimental methods to incorporate the time periodic driving field. For example, in the ultracold atom realizations of the Harper model \cite{exp0,exp01}, which make use of non-interacting Bose-Einstein condensate (BEC) under a 1D optical lattice, the time dependent term $\propto \cos(\Omega t)$ can be obtained by linearly chirping the frequencies of two counter-propagating waves \cite{Longwen, chirp}. Meanwhile, in the optical waveguide realization proposed by Ref.~\cite{exp1}, where time is simulated by the propagation distance of the light, the time dependent term $\propto \cos(\Omega t)$ can be implemented by varying the refractive index of each waveguide periodically along its length.

In order to realize the fully quantum version of our model, ultracold atom realizations of the Harper model \cite{exp0,exp01} might be more suitable as a starting point. Interaction with a quantized driving field can be simulated by placing the non-interacting BEC systems inside a quantum $LC$ circuit \cite{qexp1}. Alternatively, as proposed by Ref.~\cite{q2}, optical cavity setups can be used,  and single mode photon field can be selected from a ladder of cavity modes by using a dispersive element and dielectric mirrors. The coupling strength $V_2$ can be tuned by varying the position of the mirrors. Finally, we note that strong coupling regime between optical cavities and atomic gases or various qubit systems have been achieved experimentally \cite{qexp2,qexp3,qexp4,qexp5}. This opens up many other possibilities to realize our model.

\subsection{Towards possible detection of type-II Weyl points}\label{detect}
	
Here we discuss one possible way to manifest type-II Weyl points and distinguish them from type-I Weyl points via applying
an artificial magnetic field. It was shown recently that the tilting term in the Weyl Hamiltonian causes a ``squeezing" in the Landau level solutions if the direction of the magnetic field is perpendicular to the direction of the tilt \cite{LL, LL2}.
Under such a magnetic field, as the Weyl points undergo a transition from type-I to type-II, the Landau levels  are expected to collapse \cite{LL}, namely, the two bands in the vicinity of the type-II Weyl points start to overlap with each other.  For our CDODHM with only one physical dimension, Artificial magnetic field \cite{AMF1,AMF2,AMF3} can be introduced to simulate the effect of magnetic field in real 3D systems. For example, in order to { simulate a magnetic field along $y$ direction}, which corresponds to the vector potential $\mathcal{A}=(0,0,-Bx)$ in the Landau gauge, Peierls substitution amounts to modifying $\phi_z\rightarrow \phi_z+e B x$, so that Eq.~(\ref{Ham}) becomes,
\begin{eqnarray}
H(B) &=&\sum_n \left\lbrace\left[J+(-1)^n\lambda\cos(\phi_y)\right]\hat{c}^\dagger_{n+1}\hat{c}_n+h.c. \right\rbrace \nonumber \\
&& +\sum_n (-1)^n [V_1+V_2\cos(\Omega t)]\cos(\phi_z+eB n)\hat{c}_n^\dagger\hat{c}_n \;
\label{ll}
\end{eqnarray}

\n in the semiclassical case. It is seen above that such artificial magnetic field is achieved by a lattice-site-dependent phase modulation introduced to $\phi_z$.  In the quantum case, $\cos(\Omega t)\rightarrow \frac{a+a^\dagger}{2}$ and $H_\mathrm{field}$ as given by Eq.~(\ref{field}) is added into the Hamiltonian.

By diagonalizing the Floquet operator associated with Eq.~(\ref{ll}) numerically, the quasienergy spectrum can be obtained for the semiclassical case, which is shown in Fig.~\ref{llc}. In order to make a comparison with the fully quantum case, we are focusing on the Weyl points at quasienergy $\pi$, which may turn into type-II Weyl points in the quantum regime, and hence we choose the region of the quasienergy to be in $[0,2\pi]$. As is evident from the figure, in the vicinity of the Weyl points at quasienergy $\pi$ (Weyl points marked by the green dashed line), the Landau level structures remain qualitatively the same regardless of the value of the coupling strength $V_2$ when {the lattice-site-dependent phase modulation is added. In order to understand the robustness of the Landau level structures near the Weyl points, we calculate the quasienergies associated with Eq.~(\ref{effH}) but now under such a lattice-site-dependent phase modulation. Because here we treat an effective Hamiltonian exactly like that of a Dirac Hamiltonian in the presence of a magnetic field, we easily find

\begin{eqnarray}
\varepsilon_{n\neq 0} &=& l\pi -\mathrm{sgn}(n) \sqrt{v_0^2 k_y^2+|n|\omega_c^2} \;, \label{l1} \\
\varepsilon_{0} &=& l\pi +v_0 k_y\;, \label{l2}
\end{eqnarray}

\n where $v_0= 2\lambda J_l(lc)$ and $\omega_c =\sqrt{4eV_1 J\sin(\phi_l)J_l(lc)B}$. Eq.~(\ref{l1}) and Eq.~(\ref{l2}) imply that the quasienergy solutions are independent of $\phi_z$ (which is somewhat expected because eigenvalues of Landau levels should not depend on where electrons are). This explains the observation of plateaus in the vicinity of the Weyl points in
Fig.~\ref{llc}.  In addition, when $k_y=0$ ($\phi_y=\frac{\pi}{2}$), the zeroth Landau level quasienergy $\varepsilon_0$ is equal to an integer multiple of $\pi$. Finally, we note that the only effect of the coupling strength $V_2$ ($c \equiv V_2/V_1$) here in Eq.~(\ref{l1}) and Eq.~(\ref{l2}) is to renormalize $v_0$ and $\omega_c$ via the Bessel function $J_l(lc)$, without modifying the form of the quasienergy solutions.
}

In the fully quantum case,  the first four bands of the energy spectrum have also been obtained numerically in Fig.~{\ref{llq}. By focusing on the Weyl points along the green dotted line, which acquire a tilt as the coupling strength is tuned (i.e., these Weyl points in panel (c) have more tilting compared to those in panel (a)), it is evident that when the tilting term is not too large (the Weyl points still belong to type-I), the Landau level structures around the green dotted line in the vicinity of these Weyl points remain qualitatively the same, as shown in Fig.~\ref{llq}b.  However, as the tilting term gets larger such that a transition from type-I to type-II Weyl points takes place, these Landau level structures collapse (levels start to overlap one another), as is depicted in Fig.~\ref{llq}d around the green dotted line in the vicinity of the original type-II Weyl points. By contrast, the Weyl points along the red dotted line do not acquire any tilt as the coupling strength is varied. As a result, in both Fig.~\ref{llq}b and Fig.~\ref{llq}d, the Landau level structures around the red dotted line do not change much. { This observation can also be understood in terms of the Landau level solutions of the effective Hamiltonian near these Weyl points. Near the Weyl points marked by the red dotted line, the effective Hamiltonian takes the same form as Eq.~(\ref{effH}), thus leading to similar quasienergy solutions and properties (i.e., robustness of the quasienergy structures under a change in the phase parameter $\phi_z$ and coupling strength $V_2$) as Eq.~(\ref{l1}) and Eq.~(\ref{l2}) we have elucidated earlier. Near the Weyl points marked by the green dotted line, the technique introduced in \cite{LL} can be applied to derive the energy solutions associated with Eq.~(\ref{weyl}) under the lattice-site-dependent phase modulation introduced to $k_z$. The derivations are not trivial \cite{LL} and we finally obtain

\begin{eqnarray}
E_{n,m\neq 0} &=& \pi(2n-1)-\frac{\pi V_2^2}{8V_1^2}-\mathrm{sgn}(m)\sqrt{\alpha^2 v_0^2 k_y^2 +|m|\alpha^3 \omega_c^2}\;, \label{v3}\\
E_{n,0} &=& \pi(2n-1)-\frac{\pi V_2^2}{8V_1^2}+\alpha v_0 k_y\;, \label{v4}
\end{eqnarray}		

\n where $\alpha=\sqrt{1-\beta^2}$, $\beta=\frac{V_2}{2V_1}$, $v_0$ and $\omega_c$ are similar with those in Eq.~(\ref{l1}) and Eq.~(\ref{l2}) with $J_l(lc)$ replaced by $J_1\left(\frac{\sqrt{n} V_2}{V_1}\right)$. Due to the additional of $\alpha$ factor in Eq.~(\ref{v3}) and Eq.~(\ref{v4}), the spacing between each Landau level decreases. Moreover, for type-II Weyl points, we have $V_2>2V_1$, which implies $\beta>1$. As a result, Eq.~(\ref{v3}) and Eq.~(\ref{v4}) become imaginary and no longer correctly describes the energy structures near such Weyl points, i.e., Landau level solutions collapse \cite{LL}.		
}
	
The above observed Landau level collapse in the vicinity of type-II Weyl points suggests a possible detection of type-II Weyl points by using ideas borrowed from standard means such as the Shubnikov-de Haas oscillations or the scanning tunneling spectroscopy (STS) as mentioned in Ref.~\cite{LL}. In addition, since the generation of an artificial magnetic field only involves the modification of the phase parameter $\phi_z$, it should be feasible in terms of the experimental proposals elucidated in Sec.~{\ref{exper}}. The measurement of the {Landau level structures under the introduction of such a lattice-site-dependent phase modulation thus provides a physical way to distinguish type-II from type-I Weyl points in our physically 1D model}.

\begin{figure}
	\begin{center}
		\includegraphics[scale=0.5]{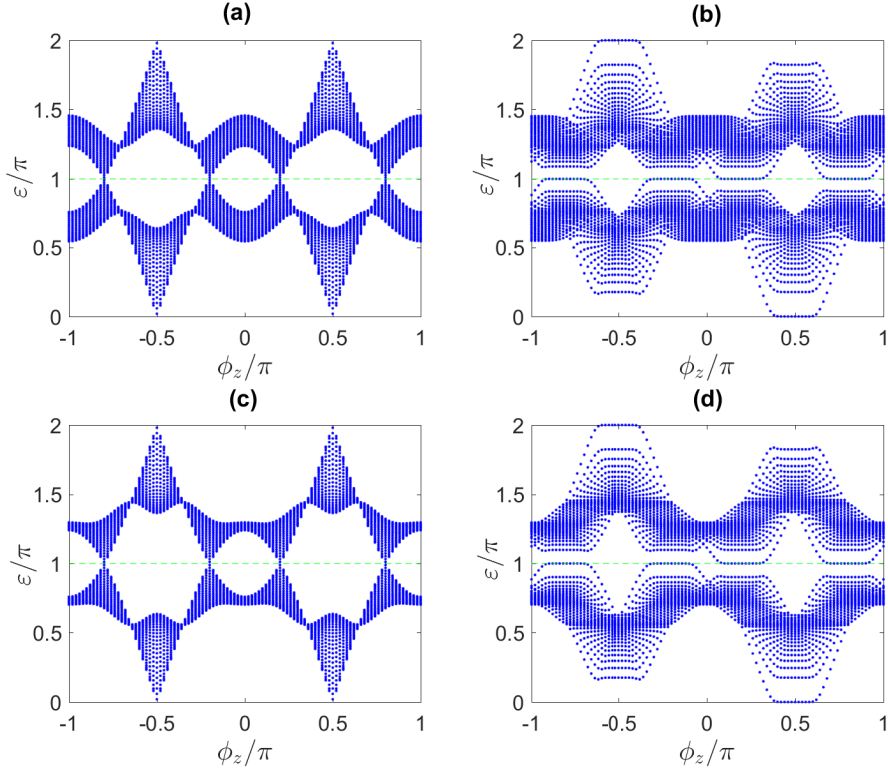}
	\end{center}
	\caption{(color online). Landau level structures around Weyl points (see green dashed line) in the semiclassical case when a { lattice-site-dependent phase modulation is introduced to $\phi_z$}.  Parameters chosen are $J=1$, $\lambda=0.5$, $\phi_y=\frac{\pi}{2}$, $V_1=\frac{\pi}{\cos(0.2\pi)}$, (a, b) $V_2=4$, (c, d) $V_2=9$. (a, c) are plotted without  lattice-site-dependent phase modulation, (b, d) are plotted under $eB=0.02$. The quasienergy region is chosen to be in $[0,2\pi]$.}
	\label{llc}
\end{figure}

\begin{figure}
	\begin{center}
		\includegraphics[scale=0.5]{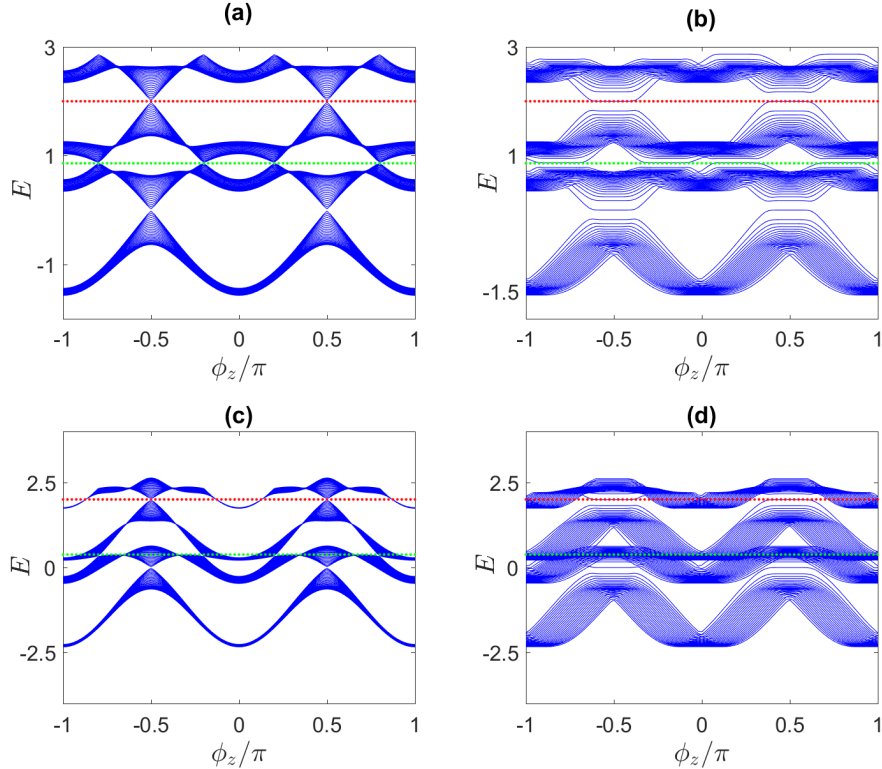}
	\end{center}
	\caption{(color online). Landau level structures around Weyl points (see red and green dotted lines) in the fully quantum case when a  lattice-site-dependent phase modulation is introduced to $\phi_z$. The parameters used are the same as those used in Fig.~(\ref{llc}).  Note that in panel (b) where only type-I Weyl points present, two bands on each side of the green or red line can still be clearly seen after the {Thursday, January 26, 2017  15:57 lattice-site-dependent phase modulation is introduced to $\phi_z$}; whereas in panel (d), the two bands around the green line (but not around the red line) start to overlap each other in the vicinity of Type-II Weyl points. }.
	\label{llq}
\end{figure}

\section{Conclusions}\label{conc}

In this paper, we consider an extension to our previous work \cite{Radit} to explore the generation of novel topological phases by using a more realistic driving term, i.e., in the form of a harmonic driving field. We then show that an interaction between the ODHM and a harmonic driving field leads to the emergence of additional Weyl points, similar to the ODKHM studied in \cite{Radit}. However, the simplicity of the model considered in this paper allows us to study the system in both full quantum (quantum field) and semiclassical (classical field) pictures.

When the driving field is treated classically as a time dependent potential, we have found using Floquet theory the locations at which new Weyl points emerge. By expanding the Floquet operator around the Weyl points, we are able to show that these Weyl points belong to type-I Weyl points. The topological signatures of these Weyl points are confirmed by the existence of Fermi arc edge states connecting each pair of Weyl points of opposite chiralities when the Floquet operator is diagonalized under OBC. Furthermore, by driving a localized Wannier state along a closed loop in parameter space, the change in its position expectation value is proportional to the total chirality of the Weyl points enclosed.

When the field is treated quantum mechanically, i.e., by taking both the atom and photons as a single system, we have shown that Weyl points emerge at the same locations as those found in the classical field case. However, some of these Weyl points acquire an extra tilting and energy shifting terms which depends on the matter-light coupling strength $V_2$. As a result, when $V_2$ is sufficiently large, it is possible for some of these type-I Weyl points to transform into type-II Weyl points. In addition, since both extra terms will not affect the Weyl points at $\left(k,\phi_y,\phi_z\right)=\left(\pm \frac{\pi}{2}, \pm \frac{\pi}{2}, \pm \frac{\pi}{2} \right)$, it is possible to generate a pair of mixed Weyl points with tunable energy difference, which opens up a possibility to realize or explore further the properties of such mixed Weyl semimetals. We have also verified that Fermi arc edge states connecting two Weyl points of opposite chiralities emerge. Moreover, via the quantization of adiabatic transport, we confirm that the chirality of the Weyl points is preserved under the transition from type-I to type-II. Possible experimental realizations have also been briefly discussed for both the semiclassical and fully quantum case. Finally, a scheme to distinguish type-II from type-I Weyl points discovered in our 1D system has also been elucidated.

Following this paper, we could now focus on studying the properties of more general Weyl semimetal systems which possess both type-I and type-II Weyl points, e.g., chiral anomaly induced transport properties, and verify them experimentally by designing an experimental realization of our model. It might also be interesting to design an experimental scheme which can realize both semiclassical and full quantum versions of our model within a single framework to observe the quantum to classical transition occurring in the model. There are some other aspects that deserve further explorations. For example, given that an interaction with a single photon mode gives rise to such controllable novel topological phases, considering multimode photon fields is imagined to be more fruitful. However, even with just a single photon mode, a possible future direction might be to consider its interaction with a topologically nontrivial many-body system (such a set up is also related to superradiant phase transition \cite{Dicke}). Finally, it is hoped that such a controllable mixed Weyl semimetal system we discovered can be useful for future devices.

\vspace{1cm}

\n {\bf Acknowledgements:} We thank Longwen Zhou for helpful discussions.

\appendix

\section{Derivation of Eq.~(\ref{Uw})} \label{app1}

Consider a rotating frame which corresponds to a transformation $|\psi\rangle \rightarrow R|\psi \rangle$, where $R=\exp\left(\mathrm{i}\frac{V_2\cos(\phi_z)\sin(\Omega t)}{\hbar \Omega}\sigma_z\right)$. The Hamiltonian in this new frame is given by

\begin{eqnarray}
\mathcal{H}_k'&=& \left[2J\cos(k) \cos(2a)+2\lambda \sin(k)\cos(\phi_y) \sin(2a)\right]\sigma_x  \nonumber \\
&&+ \left[-2J\cos(k) \sin(2a)+2\lambda \sin(k)\cos(\phi_y) \cos(2a)\right]\sigma_y +V_1 \cos(\phi_z) \sigma_z \;,
\label{rotham}
\end{eqnarray}

\n where $a=\frac{V_2\cos(\phi_z)\sin(\Omega t)}{\hbar \Omega}$. Near a band touching point at $(k,\phi_y,\phi_z)=\left(\frac{\pi}{2},\frac{\pi}{2}, \phi_l\right)$, where $\phi_l$ is as defined in the main text, Eq.~(\ref{rotham}) can be approximated as

\begin{eqnarray}
\mathcal{H}_k' &\approx& \left\lbrace-2Jk_x\cos[lc \sin(\Omega t)]-2\lambda k_y \sin[lc\sin(\Omega t)]\right\rbrace \sigma_x\nonumber \\
&& +\left\lbrace 2Jk_x\sin[lc \sin(\Omega t)]-2\lambda k_y \cos[lc\sin(\Omega t)]\right\rbrace \sigma_y +[l\pi-V_1 k_z\sin(\phi_l)]\sigma_z \nonumber \\
&=& \mathcal{H}_\mathrm{pert}+ [l\pi-V_1 k_z\sin(\phi_l)]\sigma_z \;,
\label{hamapp}
\end{eqnarray}

\n where $k_x$, $k_y$, $k_z$, and $c$ are as defined in the main text. By applying the time dependent perturbation theory, a one period time evolution operator in the interaction picture can be obtained as \cite{Sakurai},

\begin{eqnarray}
	U_I (1,0) &\approx & I-\int_0^1 \exp\left\lbrace\mathrm{i}[l\pi-V_1 k_z\sin(\phi_l)]t \right\rbrace \mathcal{H}_\mathrm{pert}  \exp\left\lbrace-\mathrm{i}[l\pi-V_1 k_z\sin(\phi_l)]t \right\rbrace dt \; \nonumber \\
	&=& I+\mathrm{i}\int_0^1 dt \left(2Jk_x\sigma_x+2\lambda k_y \sigma_y\right) \cos\left\lbrace 2[l\pi-V_1 k_z\sin(\phi_l)]t +lc\sin(\Omega t)\right\rbrace   \nonumber \\
	&& +\mathrm{i}\int_0^1 dt \left(2Jk_x\sigma_x-2\lambda k_y \sigma_y\right) \sin\left\lbrace 2[l\pi-V_1 k_z\sin(\phi_l)]t+lc\sin(\Omega t)\right\rbrace \nonumber \\
	&=& I+\mathrm{i}\left(2Jk_x\sigma_x+2\lambda k_y \sigma_y\right) J_{l-\frac{V_1 k_z \sin(\phi_l)}{2\pi}}(lc) \nonumber \\
	&\approx & I+\mathrm{i}\left(2Jk_x\sigma_x+2\lambda k_y \sigma_y\right) J_l(lc) \;.
	\label{Uint}
\end{eqnarray}

Finally, in order to obtain the Floquet operator, which is interpreted as a one period time evolution operator in the Schrodinger picture, i.e., $\mathcal{U}(k_x,k_y,k_z)=U(1,0)$, we need to convert Eq.~(\ref{Uint}) back to the Schrodinger picture. Therefore,

\begin{eqnarray}
U(1,0) &\approx & \exp\left\lbrace-\mathrm{i} [l\pi-V_1 k_z\sin(\phi_l)]\sigma_z\right\rbrace\left[I+\mathrm{i}\left(2Jk_x\sigma_x+2\lambda k_y \sigma_y\right) J_l(lc)\right] \nonumber \\
&\approx & \exp\left(-\mathrm{i}l\pi\right)\left[I+\mathrm{i} V_1 k_z\sin(\phi_l)\sigma_z\right]\left[I+\mathrm{i}\left(2Jk_x\sigma_x+2\lambda k_y \sigma_y\right) J_l(lc)\right] \nonumber \\
&\approx & \exp\left(-\mathrm{i}l\pi\right)\left[I+\mathrm{i} V_1 k_z\sin(\phi_l)\sigma_z+\mathrm{i}\left(2Jk_x\sigma_x+2\lambda k_y \sigma_y\right) J_l(lc)\right] \nonumber \\
&\approx & \exp\left\lbrace-\mathrm{i}\left\lbrace l\pi-\left[ V_1 k_z\sin(\phi_l)\sigma_z+2Jk_x J_l(l c)\sigma_x +2\lambda k_y J_l(l c)\sigma_y\right]\right\rbrace\right\rbrace \;,
\end{eqnarray}

\n which proves Eq.~(\ref{Uw}).

\section{Derivation of Eq.~(\ref{weyl})} \label{app2}

We start by introducing the following unit vectors,

\begin{eqnarray}
	\hat{n}&=&\frac{2J\cos(k)\hat{x}+2\lambda\sin(k)\cos(\phi_y)\hat{y}+V_1\cos(\phi_z)\hat{z}}{\frac{1}{2} \omega}\; , \\
	\hat{m} &=& \frac{-2\lambda \sin(k) \cos(\phi_y) \hat{x}+2J\cos(k)\hat{y}}{\frac{1}{2}\omega'}\; , \\
	\hat{l} &=& -\frac{\omega'}{\omega}\hat{z}+V_1\cos(\phi_z)\frac{2\lambda\sin(k)\cos(\phi_y)\hat{y}+2J\cos(k)\hat{x}}{\frac{1}{4} \omega \omega'} \; ,
\end{eqnarray}

\n where $\hat{x}$, $\hat{y}$, and $\hat{z}$ are unit vectors along $x$, $y$, and $z$ direction, $\frac{1}{2}\omega =\sqrt{\frac{1}{4}\omega'^2+V_1^2\cos^2(\phi_z)}$ and $\frac{1}{2}\omega' =\sqrt{4J^2\cos^2(k)+4\lambda^2\sin^2(k)\cos^2(\phi_y)}$. It can be verified that $\hat{l}$, $\hat{m}$, and $\hat{n}$ are three unit vectors that form a right-handed system similar to $\hat{x}$, $\hat{y}$, and $\hat{z}$. Next, we define $\sigma_\pm =\hat{l}\cdot \sigma \pm \mathrm{i} \hat{m}\cdot \sigma$. If $|\psi_\pm \rangle$ is the eigenstate of $\hat{n}\cdot \sigma$ corresponding to eigenvalue $\pm 1$, then $\sigma_+ |\psi_+ \rangle = \sigma_- |\psi_- \rangle = 0$, $\sigma_+ |\psi_-\rangle =2 c_+ |\psi_+ \rangle$ and $\sigma_- |\psi_+\rangle =2 c_- |\psi_- \rangle$, where $c_\pm$ is a unit complex numbers which depends on the representation of the eigenstates. It can also be shown that $\sigma_\pm $ and $\hat{n}\cdot \sigma$ satisfy the following algebra,

\begin{eqnarray}
	[\sigma_-, \sigma_+] &=& -4\hat{n}\cdot \sigma \;, \\
	\left[ \sigma_\pm , \hat{n}\cdot \sigma \right] &=& \mp 2\sigma_\pm \; .
\end{eqnarray}

In terms of the notations defined above, Eq.~(\ref{Htotk}) can be recast in the following form,

\begin{eqnarray}
	\mathcal{H}_q &=& \frac{1}{2}\omega \hat{n}\cdot \sigma +\Omega a^\dagger a -\frac{V_2\cos(\phi_z)\omega'}{4\omega}(a+a^\dagger)\left[\sigma_+ + \sigma_- -\frac{4V_1\cos(\phi_z)}{\omega'} \hat{n}\cdot \sigma \right] .
	\label{Quant4}
\end{eqnarray}

\n In $X$ representation and in the basis $\left\lbrace |\psi_+\rangle , |\psi_-\rangle \right\rbrace$, where $X$ is one of the quadrature operators defined in the main text, the energy eigenvalue equation associated with Eq.~(\ref{Quant4}) near $(k,\phi_y,\phi_z)=\left(\frac{\pi}{2}, \frac{\pi}{2},\phi_1 \right)$, up to first order in $k_x$, $k_y$, and $k_z$ defined in the main text, can be written as

\begin{equation}
	\left(\begin{array}{cc} A(x)+B(x) & C(x)\omega' c_- \\ C(x) \omega' c_+ & A(x)-B(x) \end{array}\right) \left(\begin{array}{c}f_1(x) \\ f_2(x) \end{array}\right) = E \left(\begin{array}{c}f_1(x) \\ f_2(x) \end{array}\right) \;,
	\label{eiv}
\end{equation}

\n where $x$ is the eigenvalue of $X$, $E$ is the energy eigenvalue, and

\begin{eqnarray}
	A(x) &=& \frac{1}{2}\Omega \left(x^2 -\frac{\partial^2}{\partial x^2}-1\right)\;, \\
	B(x)&\approx & \frac{1}{2}\omega +\frac{V_1V_2\left[\frac{\pi^2}{V_1^2}+\frac{2\pi k_z}{V_1}\sin(\phi_1)\right]}{\omega} \sqrt{2} x \;,  \\
	C(x) &\approx & -\frac{V_2}{4V_1} \sqrt{2} x\;.
\end{eqnarray}

Since $k_x$ and $k_y$ are very small quantities, $\omega'$ is also very small by construction and thus the off-diagonal terms in Eq.~(\ref{eiv}) can be treated as perturbations. Without the off-diagonal terms, Eq.~(\ref{eiv}) reduces to two uncoupled harmonic oscillator eigenvalue equations, which can readily be solved for the energy $E^{(0)}$ and the eigenfunctions $f_1(x)$ and $f_2(x)$. In particular,

\begin{equation}
	E_{n,\pm}^{(0)} =  \pi (2n\pm 1) \mp V_1 k_z\sin(\phi_1) -\frac{\pi V_2^2}{8V_1^2} +\frac{V_2^2}{4V_1}k_z \sin(\phi_1)\;,
	\label{unperbE}
\end{equation}

\n where $n$ is a non-negative integer.

\begin{figure}
	\begin{center}
		\includegraphics[scale=0.45]{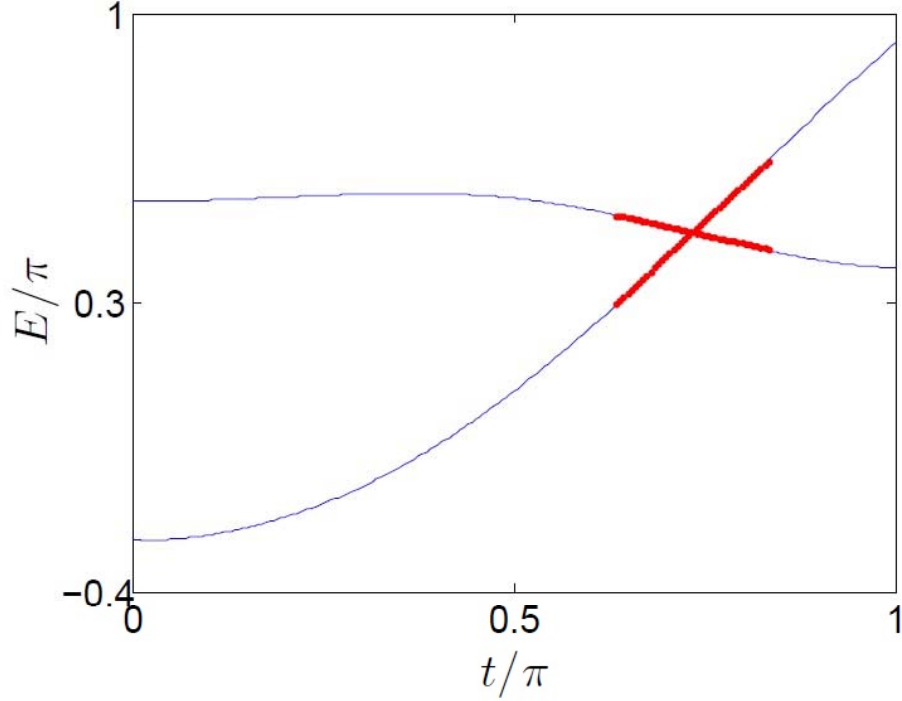}
	\end{center}
	\caption{(color online). A typical band structure of the quantized CDODHM along a curve in the Brillouin zone parameterized by $t$ according to $(k(t),\phi_y(t),\phi_z(t))=(\frac{0.5 \pi t}{r}, \frac{0.5 \pi t}{r}, \frac{0.2 \pi t}{r})$, where $r=\sqrt{0.5^2+0.5^2+0.2^2}$. The blue curve is obtained by diagonalizing Eq.~(\ref{Quant4}), whereas the red dotted curve is obtained by plotting Eq.~(\ref{weyl2}) near a Weyl point. The system parameters are $J=1$, $\lambda=0.5$, $V_1=\frac{\pi}{\cos\left(0.2\pi\right)}$, and $V_2=8$.}
	\label{compare}
\end{figure}

To understand the effect of the off-diagonal term, we define the following operators,

\begin{eqnarray}
\mathcal{A}_+ &=& \frac{1}{\sqrt{2}}\left[\left(X+\frac{\sqrt{2}V_2}{2V_1}\right)+\mathrm{i} P\right]\;, \\
\mathcal{A}_- &=&\frac{1}{\sqrt{2}}\left[\left(X-\frac{\sqrt{2}V_2}{2V_1}\right)+\mathrm{i} P\right]\;.
\end{eqnarray}

\n The off-diagonal perturbation term and the unperturbed diagonal term can then be written as, respectively,

\begin{eqnarray}
H_\mathrm{off} &=& -\frac{V_2}{8V_1}\omega' \left(\mathcal{A}_+^\dagger +\mathcal{A}_-\right)\left(\sigma_+ +\sigma_-\right)\;, \label{pert} \\
H_\mathrm{on} &=& \left[\pi -V_1\sin(\phi_1)k_z\right] \left(\begin{array}{cc} 1 & 0 \\ 0 & -1 \end{array}\right) -\frac{\pi V_2^2}{8V_1^2}+\frac{V_2^2}{4V_1}k_z \sin(\phi_1)+2\pi \left(\begin{array}{cc} \mathcal{A}_+^\dagger \mathcal{A}_+ & 0 \\ 0 & \mathcal{A}_-^\dagger \mathcal{A}_- \end{array}\right) \;. \label{unpert} \nonumber \\
\end{eqnarray}

\n Since $\omega'$ is a very small quantity, rotating wave approximation (RWA) could be made if $\mathcal{A}_+^\dagger$ and $\sigma_-$, as well as $\mathcal{A}_-$ and $\sigma_+$, are governed by approximately the same frequency of evolution. Therefore, let's first analyze the equations of motion for $\mathcal{A}_\pm$ and $\sigma_\pm$ (in the interaction picture):

\begin{eqnarray}
\frac{d\sigma_\pm}{dt} &=& \frac{[\sigma_\pm, H_\mathrm{on}]}{\mathrm{i}} \nonumber \\
&\approx& \frac{\mp 2\pi \sigma_\pm}{\mathrm{i}}\mp \frac{2V_2\pi}{\mathrm{i} V_1} \left(\mathcal{A}_+^\dagger +\mathcal{A}_-\right)\sigma_\pm \;, \label{evol1}\\
\frac{d\mathcal{A}_\pm}{dt} &=& \frac{[\mathcal{A}_\pm, H_\mathrm{on}]}{\mathrm{i}}  \nonumber \\
&=& \frac{2\pi \mathcal{A}_\pm}{\mathrm{i}}\mp\frac{\pi V_2}{\mathrm{i} V_1}+\frac{\pi V_2}{\mathrm{i} V_1} \left(\begin{array}{cc} 1 & 0 \\ 0 & -1 \end{array}\right) \;.
\label{evol2}
\end{eqnarray}

Let's first assume $V_2$ to be sufficiently small, so that the solutions to the above equations are approximately $\sigma_\pm (t) \approx \sigma_\pm (0) e^{\pm \mathrm{i} 2\pi t}$ and $\mathcal{A}_\pm (t) \approx \mathcal{A}_\pm (0) e^{- \mathrm{i} 2\pi t}$. RWA can then be invoked, and the total Hamiltonian can be divided into subblocks spanned by the states $|n,-\rangle$ and $|n-1,+\rangle$, which are eigenstates of $H_\mathrm{on}$ corresponding to $E_{n,-}^{(0)}$ and $E_{n-1,+}^{(0)}$ as given in Eq.~(\ref{unperbE}) respectively. The reduced $2\times 2$ Hamiltonian in $\left\lbrace|n,-\rangle, |n-1,+\rangle\right\rbrace$ basis is,

\begin{eqnarray}
	{[H_q]_n} &\hat{=}& \left(\begin{array}{cc} E_{n-1,+}^{(0)} & -\frac{V_2}{4V_1}\omega' \sqrt{n} c_+ \\ -\frac{V_2}{4V_1}\omega' \sqrt{n} c_- & E_{n,-}^{(0)} \end{array}\right) \;.
\end{eqnarray}

\n By considering a representation where $c_-=\frac{4Jk_x}{\omega'}+\mathrm{i} \frac{4\lambda k_y}{\omega'}$, and $\tau_x$, $\tau_y$, and $\tau_z$ take the usual Pauli matrices form, the reduced Hamiltonian can be written more compactly as,

\begin{eqnarray}
{[H_q]_n} &=& \pi (2n-1)-\frac{\pi V_2^2}{8V_1^2} +\frac{V_2^2}{4V_1}k_z \sin(\phi_1)-V_1 \sin(\phi_1) k_z\tau_z -\left(2J k_x \tau_x + 2\lambda k_y \tau_y\right) \frac{\sqrt{n}V_2}{2V_1}\;. \nonumber \\
\label{almost}
\end{eqnarray}

Let's now relax the assumption that $V_2$ is sufficiently small. We notice that $\frac{\sqrt{n}V_2}{2V_1}$ corresponds to the lowest order term in the series expansion of a certain function, e.g. $J_1\left(\frac{\sqrt{n}V_2}{V_1}\right)$. Since the Hamiltonian is required to reduce to Eq.~(\ref{effH}) in the classical limit $n\rightarrow \infty$ and $V_2\rightarrow 0$, we argue that for an arbitrary value of $V_2$ (not necessarily small), Eq.~(\ref{almost}) need to be modified by replacing the $\frac{\sqrt{n}V_2}{2V_1}$ factor in the $\tau_x$ and $\tau_y$ terms by $J_1\left(\frac{\sqrt{n}V_2}{V_1}\right)$, so that Eq.~(\ref{weyl}) follows. Although this argument is not obvious to be justified analytically, it is still possible to verify Eq.~(\ref{weyl}) numerically by comparing the eigenvalues of Eq.~(\ref{Quant4}) obtained directly from exact diagonalization and the eigenvalues of Eq.~(\ref{weyl}) near a Weyl point when $V_2$ is at the same order as the other parameters, as confirmed in Fig.~\ref{compare}.

%%%%%%%%%%%%%%%%%%%%%%%%%%%%%% REFERENCE %%%%%%%%%%%%%%%%%%%%%%%%%%%%%%%%%%%%%%%

\end{document}